\documentclass[]{spie}  

\usepackage{amsmath,amsfonts,amssymb}
\usepackage{graphicx}
\usepackage[colorlinks=true, allcolors=blue]{hyperref}

\usepackage{latexsym}

\usepackage{color, soul}
\usepackage{xargs}
\usepackage{tikz}
\usepackage{todonotes}
\usepackage{multirow} 
\usepackage{verbatim}
\usepackage{nameref,hyperref}

\usepackage{multicol}
\usepackage{multirow}
\usepackage{hyperref}
\usepackage{subcaption}
\usepackage{bm}
\usepackage{booktabs}

\newlength\savewidth\newcommand\shline{\noalign{\global\savewidth\arrayrulewidth
  \global\arrayrulewidth 1.5pt}\hline\noalign{\global\arrayrulewidth\savewidth}}

\title{Quantifying uncertainty in lung cancer segmentation with foundation models applied to mixed-domain datasets}

\author[a]{Aneesh Rangnekar}
\author[a]{Nishant Nadkarni}
\author[a]{Jue Jiang}
\author[a]{Harini Veeraraghavan}
\affil[a]{Department of Medical Physics, Memorial Sloan Kettering Cancer Center, NY, USA}

\authorinfo{Further author information: Send correspondence to A. Rangnekar at \href{mailto:rangnea@mskcc.org}{rangnea@mskcc.org}}

\begin{document} 
\maketitle

\begin{abstract}
Medical image foundation models have shown the ability to segment organs and tumors with minimal fine-tuning. These models are typically evaluated on task-specific in-distribution (ID) datasets. However, reliable performance on ID datasets does not guarantee robust generalization on out-of-distribution (OOD) datasets. Importantly, once deployed for clinical use, it is impractical to have `ground truth' delineations to assess ongoing performance drifts, especially when images fall into the OOD category due to different imaging protocols. Hence, we introduced a comprehensive set of computationally fast metrics to evaluate the performance of multiple foundation models (Swin UNETR, SimMIM, iBOT, SMIT) trained with self-supervised learning (SSL). All models were fine-tuned on identical datasets for lung tumor segmentation from computed tomography (CT) scans. SimMIM, iBOT, and SMIT used identical architecture, pretraining, and fine-tuning datasets to assess performance variations with the choice of pretext tasks used in SSL. The evaluation was performed on two public lung cancer datasets (LRAD: n = 140, 5Rater: n = 21) with different image acquisitions and tumor stages compared to training data (n = 317 public resource with stage III-IV lung cancers) and a public non-cancer dataset containing volumetric CT scans of patients with pulmonary embolism (n = 120). All models produced similarly accurate tumor segmentation on the lung cancer testing datasets. SMIT produced the highest F1-score (LRAD: 0.60, 5Rater: 0.64) and lowest entropy (LRAD: 0.06, 5Rater: 0.12), indicating higher tumor detection rate and confident segmentations. In the OOD dataset, SMIT misdetected the least number of tumors, marked by a median volume occupancy of 5.67 cc compared to the best method SimMIM of 9.97 cc. Our analysis shows that additional metrics such as entropy and volume occupancy may help better understand model performance on mixed domain datasets.

\end{abstract}


\section{Purpose/Motivation}
\label{sec:intro}  

Foundation models are developed with the goal of universal applicability to a variety of problems with minimal fine-tuning to new tasks. Once fine-tuned, these models are evaluated on selected in-domain datasets with a similar distribution as the fine-tuning datasets, using task-specific accuracy metrics. However, task-specific metrics may not be sufficient to ensure reliably accurate performance when such models are to be deployed for multi-institutional research or routine clinical use, where datasets can exhibit wide variations from training data. Performance drift occurs when test image distribution diverges from the training distribution, a challenge known as distribution shift. Another source of poor performance is concept drift when the datasets used in broader testing have diseases that differ from those used in training data (e.g. data trained on advanced-stage malignant cancers applied to segment precancerous or early-stage cancers or on non-cancer datasets. Models must be resilient to image distribution shifts and concept drifts.

Currently, a segmentation model's performance is quantified only using task-specific metrics, which provides a limited view of robustness in out-of-domain datasets. Out-of-distribution (OOD) analysis is routinely done in computer vision applications that systematically compute OOD performance metrics in addition to task-specific accuracy metrics \cite{yang2024generalized}; OOD metrics are not often used for analysis in medical image segmentations \cite{graham2022transformer,yuan2023devil}. Hence, we extended OOD metrics used in natural image analysis to evaluate OOD performance and also quantify uncertainty in the medical image segmentation applicable to cases with distribution and concept shifts. We systematically evaluated four different foundation models for generating volumetric lung tumor segmentation from contrast and non-contrast 3D CTs. Lung tumor segmentation was selected as this is more complex than organ segmentation and multiple public benchmarking CT thoracic datasets are available to assess accuracy with both distribution and concept drifts.

\section{Methods}

\textbf{Foundation models: } Swin UNETR \cite{tang2022self,Willemink2022}, SimMIM \cite{xie2022simmim}, iBOT \cite{zhou2021ibot} and SMIT \cite{jiang2022self}, trained on the 3D-Swin transformer backbone to use a consistent network backbone, were used in this study. Also, SimMIM, iBOT, and SMIT used identical encoder and decoder architecture, consistent data augmentations, and standardized training hyperparameters. Our purpose for maintaining architecture similarity was to answer: \textit{how do pretraining tasks affect downstream performance and can we quantify performance differences?}

SimMIM uses a masked image pretask that applies a random mask to the input (in our case, a 3D mask to a 3D cube) and then trains the neural network to reconstruct the masked regions. This approach encourages the model to learn contextual information as it must rely on the visible parts to predict the missing regions. iBOT extends SimMIM by learning good feature representations from masked regions via token distillation in a teacher-student framework. iBOT uses a combination of local (voxel-level) and global (scan-level) tokens for pretraining the network. SMIT combines the masked image modeling and token distillation used in SimMIM and iBOT to learn dense pixel dependencies with iBOT's local-global connections. For all our models, we used a Swin transformer with a depth of $2-2-12-2$ and an input size of $128 \times 128 \times 128$. We used the decoder heads as mentioned in the corresponding literature for pretraining and used a consistent U-Net decoder initialized with random weights for fine-tuning. Swin UNETR \cite{tang2022self} model pretrained weights and model structure (depth of $2-2-2-2$ and input size of $96 \times 96 \times 96$) were used to maintain consistency with the published work.

\begin{figure}[t]
    \centering
    \begin{subfigure}[t]{0.49\textwidth}
        \centering
        \includegraphics[width=\linewidth]{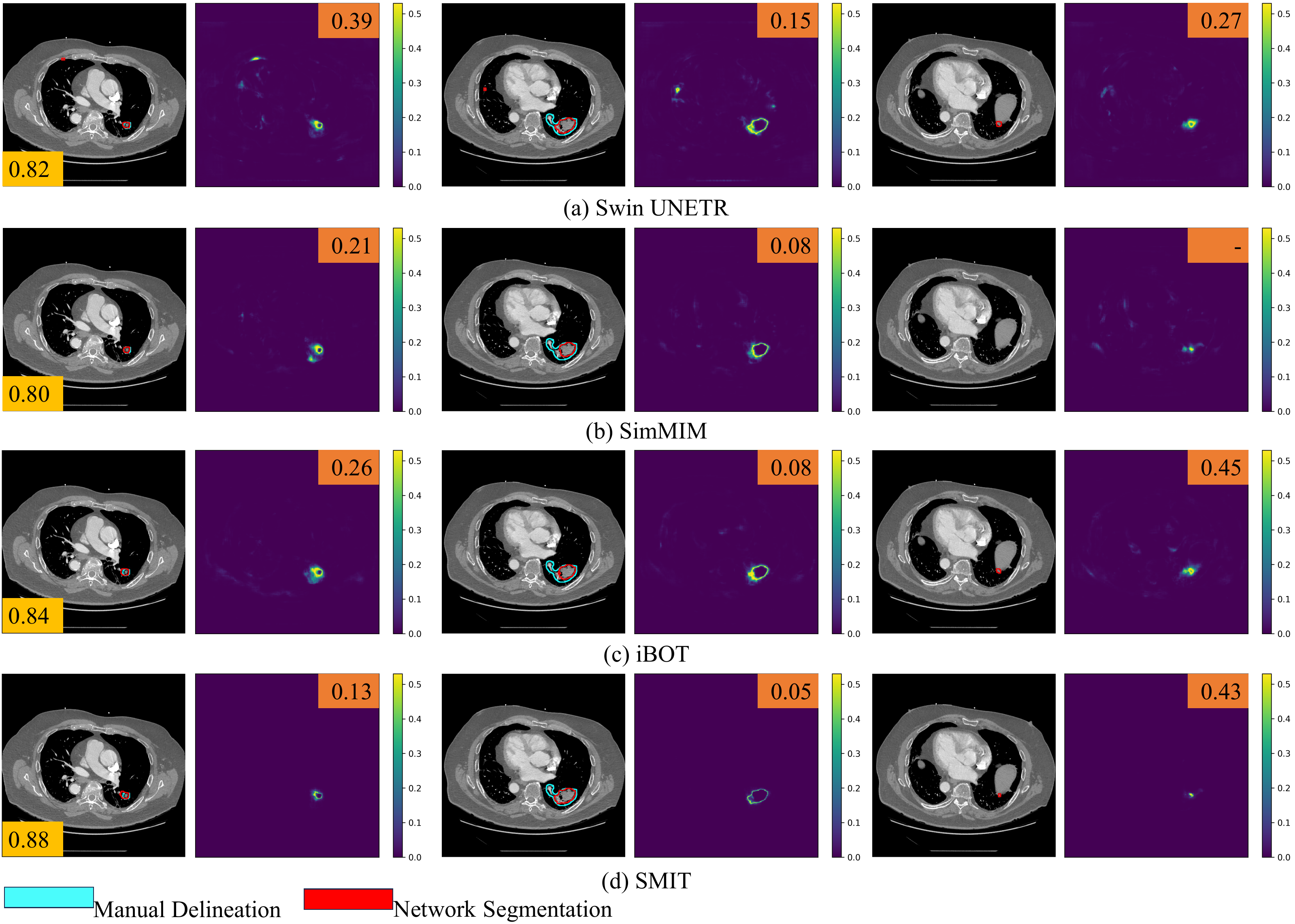}
        \caption{LRAD Visualizations}
        \label{fig:segent:lrad}
    \end{subfigure}%
    \hfill
    \begin{subfigure}[t]{0.49\textwidth}
        \centering
        \includegraphics[width=\linewidth]{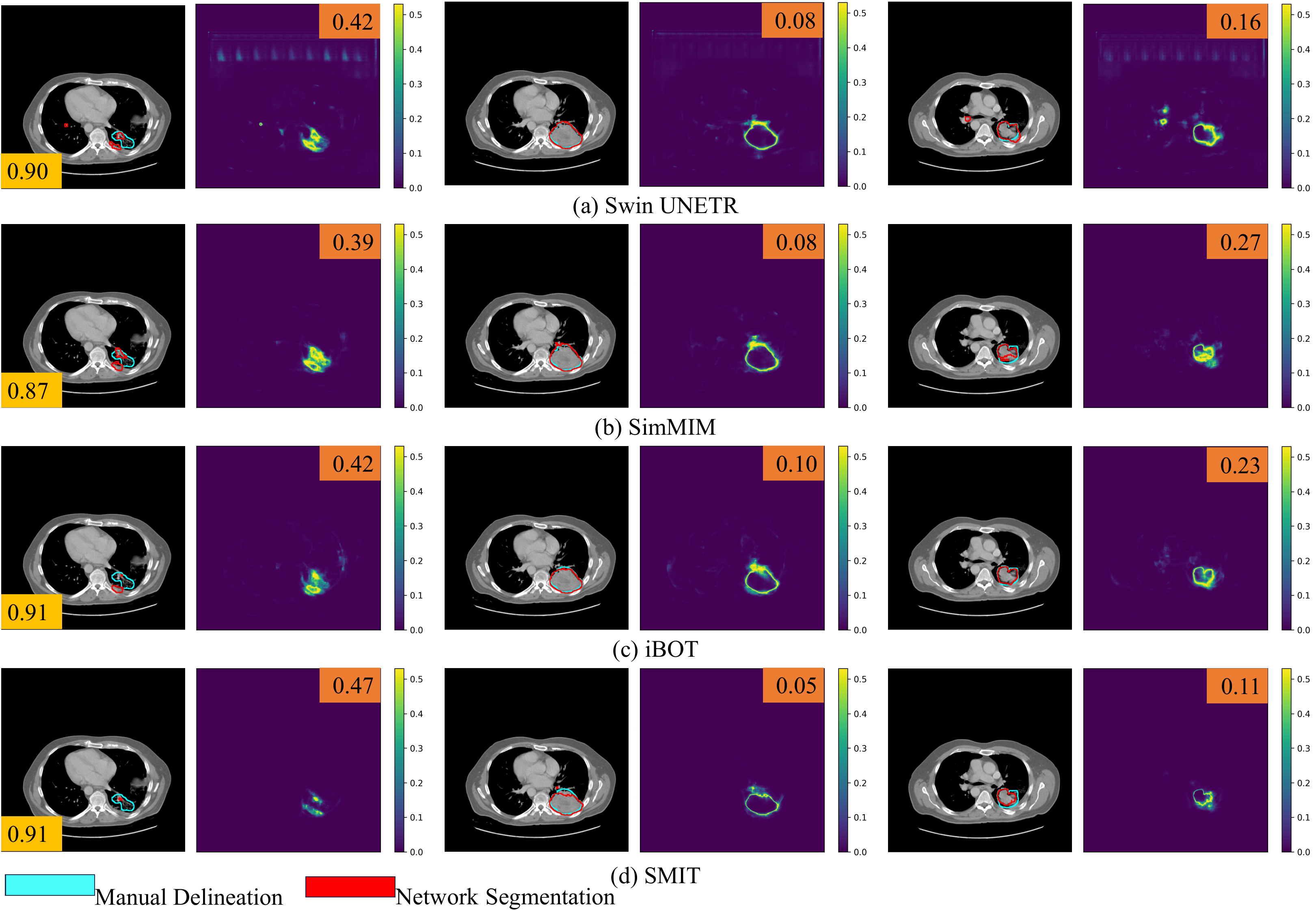}
        \caption{5Rater Visualizations}
        \label{fig:segent:5rater}
    \end{subfigure}
    \caption{Example segmentations and voxel-wise entropy shown for superior, central, and inferior slices. The average entropy for individual slices is shown on top-right. Volumetric DSC for the entire image is shown in yellow on the bottom-left. The `-' indicates that the model did not detect/segment any tumor.}
\end{figure}

\noindent\textbf{Pretraining datasets: \/}\rm We used 10,412 3D CT scans encompassing diseases from head and neck to the pelvis sourced from datasets provided publicly for a variety of tasks including lesion detection~\cite{xiao2023lesion}, classification~\cite{harmon2020artificial}, and multi-organ and abdominal tumor segmentations for pretraining. Anonymized institutional 3D-CT  datasets from patients treated for lung, esophageal, and head and neck cancers with radiation were used as is without any curation. \textbf{Fine-tuning \/}\rm was performed using a public dataset containing stage III-IV non-small cell lung cancers (NSCLC) (n = 317 3D-CTs), with a 70\%-30\% stratified training and validation split \cite{aerts2015data}. \textbf{Testing datasets: \/}\rm Two in-distribution (ID) with the same concept but different CT acquisitions (distribution shift) were evaluated. This included patients with early-stage I-II NSCLC \cite{bakr2017data} (referred to as LRAD) and stage II-IV with multi-rater segmentations\cite{Wee2019} (referred to as 5Rater). Last, one OOD dataset (concept drift) consisting of a subset of 120 annotated images from the RSNA Pulmonary Embolism (PEmb) dataset \cite{colak2021rsna,callejas2023augmentation} was evaluated to assess robustness to false detections. The OOD dataset presents an additional challenge as the disease occurs in the same anatomical region (chest CT) as used for training and is confirmed to be devoid of cancers. 

\noindent \textbf{Implementation details: \/}\rm Models were fine-tuned using Dice and Cross-entropy losses with a batch size of 16 (4 $\times$ 4 NVIDIA A100 GPUs), learning rate of $2e^{-4}$, and linear warmup scheduler with cosine annealing for 1000 epochs (with Pytorch \cite{paszke2019pytorch}). The lung window of $[-400, 400]$ HU and $1mm \times 1mm \times 1mm$ voxel spacing was used throughout all experiments. Sliding window inference with 50\% Gaussian overlap to segment volumetric CTs with varying fields of view \cite{tang2022self,jiang2022self}.

\begin{table}[t]
\centering
\def\arraystretch{1.25}
\scriptsize
\caption{Fine-tuned foundation model performances on the ID datasets. The entropy values for the values for validation set are as follows: Swin UNETR: 0.17 $\pm$ 0.06, SimMIM: 0.15 $\pm$ 0.06, iBOT: 0.17 $\pm$ 0.06, SMIT: 0.07 $\pm$ 0.03.}
\label{tab:results_id}
\resizebox{\textwidth}{!}{%
\begin{tabular}{l|llll|llll}
\multirow{2}{*}{\begin{tabular}[c]{@{}l@{}}Pretraining\\ Strategy\end{tabular}} &  \multicolumn{4}{l}{LRAD Dataset (N = 140)}  & \multicolumn{4}{l}{5Rater Dataset (N = 21)}\\
 & DSC & HD95 (mm) & F1-Score & Entropy & DSC & HD95 (mm) & F1-Score & Entropy \\ \shline
Swin UNETR \cite{tang2022self} & 0.78 $\pm$ 0.09 & 6.71 $\pm$ 7.89 & 0.481 &0.20 $\pm$ 0.07  & 0.86 $\pm$ 0.01 & 5.10 $\pm$ 7.77 & 0.416 & 0.19 $\pm$ 0.05\\
SimMIM \cite{xie2022simmim} & 0.80 $\pm$ 0.07 & 5.09 $\pm$ 4.61 & 0.553 & 0.16 $\pm$ 0.06 & 0.83 $\pm$ 0.14 & 4.65 $\pm$ 7.59 & 0.466 & 0.21 $\pm$ 0.06\\
iBOT \cite{zhou2021ibot} & 0.80 $\pm$ 0.08 & 5.53 $\pm$ 5.24 & 0.555 & 0.16 $\pm$ 0.07 & 0.87 $\pm$ 0.06 & 3.35 $\pm$ 4.00 & 0.353 & 0.23 $\pm$ 0.06 \\
SMIT \cite{jiang2022self} & 0.81 $\pm$ 0.07 & 5.39 $\pm$ 5.07 & 0.601 & 0.06 $\pm$ 0.04 & 0.86 $\pm$ 0.06 & 4.43 $\pm$ 5.51 & 0.641 & 0.12 $\pm$ 0.06 \\ \bottomrule
\end{tabular}%
}
\end{table}

\noindent{\textbf{Segmentation metrics: \/}\rm
Dice similarity coefficient (\textbf{DSC})  and Hausdorff distance at 95th percentile (\textbf{HD95}) computes the volumetric dice and distance for clustered tumor regions where the 3D overlap between manual delineations and predicted segmentation is $\geq$ 50$\%$. \textbf{F1-score} was computed as a combination of precision and recall to measure the total number of clustered regions that were incorrectly identified as tumors or completely missed by the algorithm, respectively.

\begin{figure}
    \centering
    \includegraphics[width=0.95\linewidth]{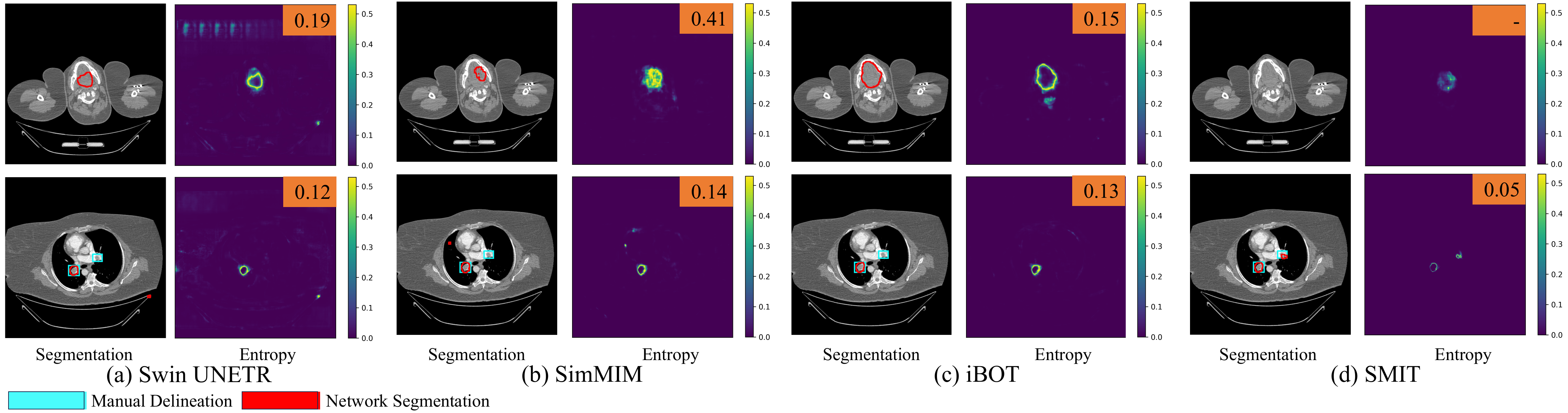}
    \caption{Example segmentations and corresponding entropy for the four models on two different images from the Pulmonary Embolism dataset. The slice-wise entropy is displayed in orange on the top-right.  The `-' indicates that the model did not detect/segment any tumor.}
    \label{fig:segent:pemb}
\end{figure}

\noindent\textbf{OOD metrics: \/}\rm Tumor segmentation can be considered a binary voxel-wise classification problem. Voxels that were classified as tumor (class $=$ 1) were used in the computation and averaged to extract the following image-level OOD metrics: \textbf{AUROC} or area under the receiver operating curve,  measures the accuracy of correctly identifying an image as an example from the OOD dataset. \textbf{FPR$@$95} represents the false positive rate of the binary classifier at the 95\% decision threshold, wherein it measures how often a model mistakenly labels an ID image as OOD when the model correctly identifies 95\% of the OOD examples \cite{hendrycks2020pretrained}. Effectively, it rates the confidence of the model predictions. \textbf{Entropy} quantifies the uncertainty in the model's tumor predictions, providing a different insight into the confidence of the model about its classification. Higher entropy indicates that the model has encountered data from an unfamiliar distribution and is less certain of its predictions. \textbf{Volume occupancy} measures the total misdetected tumor volume (in cc). This metric quantifies the actual number of voxels that were incorrectly segmented instead of averaging metrics across an entire image.

\section{Results}

\begin{figure}[t]
    \centering
    \includegraphics[width=0.9\linewidth]{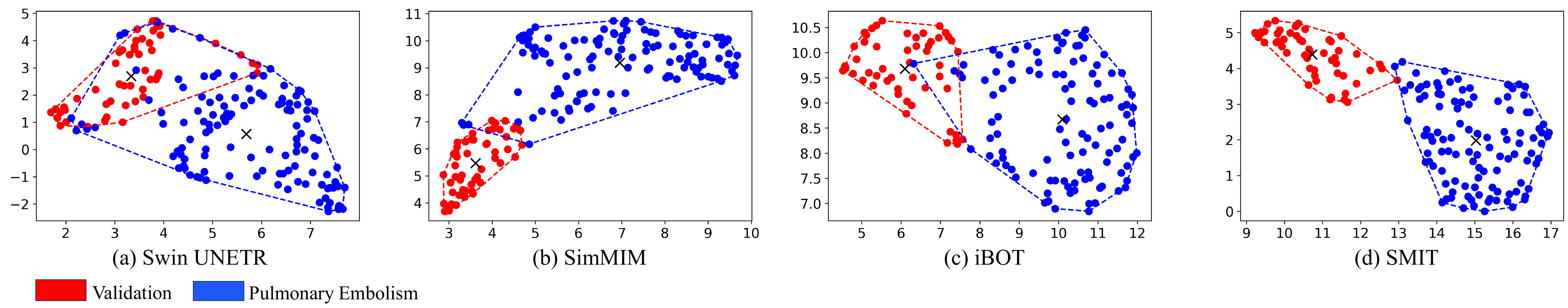}
    \caption{UMAP clustering for features of the validation and Pulmonary embolism images derived from Stage 1 of the Swin transformer architecture for all four foundation models.}
    \label{fig:clusters2}
\end{figure}

\begin{table}[t]
\centering
\def\arraystretch{1.25}
\scriptsize
\caption{OOD Detection performance of the foundation models post fine-tuning. We report results using Maximum Softmax Probability for AUROC and FPR95 and report the Volume Occupancy as median and inter-quartile range.}
\label{tab:results_ood}
\resizebox{0.7\textwidth}{!}{%
\begin{tabular}{lllll}
\multirow{2}{*}{\begin{tabular}[c]{@{}l@{}}Pretraining \\ Strategy\end{tabular}} & \multicolumn{4}{l}{Pulmonary Embolism Dataset (N = 120)} \\
 & Volume Occupancy (cc) & Entropy & AUROC & FPR@95\\ \shline
Swin UNETR \cite{tang2022self} & 16.24 [4.56 $\pm$ 75.58] & 0.27 $\pm$ 0.05 & 90.77 & 26.92 \\
SimMIM \cite{xie2022simmim} & 9.97 [1.03 $\pm$ 55.71] & 0.26 $\pm$ 0.07 & 91.65 & 23.53 \\
iBOT \cite{zhou2021ibot} & 8.46 [0.55 $\pm$ 98.24] & 0.27 $\pm$ 0.08 & 88.12 & 42.31 \\
SMIT \cite{jiang2022self} & 5.67 [0.56 $\pm$ 37.75] & 0.15 $\pm$ 0.06 & 89.58 & 34.62 \\ \bottomrule
\end{tabular}%
}
\end{table}

As shown in Table \ref{tab:results_id}, all models produced similarly accurate tumor segmentations in both datasets. However, SMIT produced a higher F1 score, indicating a higher precision and recall rate. It was also more confident compared to other methods as shown by the lowest average entropy. Of the compared methods, Swin UNETR was the least accurate and had the highest entropy. Figures \ref{fig:segent:lrad} and \ref{fig:segent:5rater} show segmentation and voxel-wise entropy for representative examples produced by all methods for superior, central, and inferior tumor extent slices. SMIT consistently had higher uncertainty at the boundary compared to other models that also showed higher entropy inside the tumor. Entropy is expected to be higher at the boundary due to the tumor and healthy tissue interface. 

Table \ref{tab:results_ood} presents the results for OOD detection. As shown, SimMIM had the best OOD performance in terms of AUROC and FPR@95 metrics. SMIT had slightly lower AUROC and FP@95 compared to SimMIM but lower entropy, indicating that this network produced confident incorrect predictions. On the other hand, SMIT also resulted in the least number of voxels being misclassified as tumors compared to all other methods. Figure \ref{fig:segent:pemb} shows that all models except SMIT falsely segment a tumor in the oral cavity in the first case. SMIT did not generate a segmentation, although uncertainty that is lower than all the other models is indicated. In the second case (bottom row), SMIT segmented both the embolisms identified by radiologists. Swin UNETR, SimMIM, and iBOT identified one of the two embolisms and generated false detections (table for Swin UNETR, top right lung region for SimMIM). Detecting pulmonary embolisms as tumors presents a dual challenge of zero-shot detection as well as conservative prediction, both of which are beyond the current scope of this paper.

To further analyze the contribution of features towards OOD performance, we computed an unsupervised clustering of stage I features extracted by the transformer networks \cite{masarczyk2024tunnel}. Features were extracted by randomly sampling two 128 $\times$ 128 $\times$ 128 3D crops containing the segmented tumors (centered on the tumor wherever applicable), taking an average representation of the voxels and then applying UMAP with $n\_neighbors$ = 20, $min\_dist$ = 0.1, and the Euclidean metric for distance. UMAP\cite{McInnes2018} cluster results are shown in Figure \ref{fig:clusters2}. As shown, there was a lot of overlap between the tumor and pulmonary embolism features for Swin UNETR and iBOT networks. SMIT and SimMIM were able to better differentiate the two datasets, indicating that feature representations extracted by these networks are more robust to OOD variations. 

\section{Conclusion}
We performed a comprehensive evaluation of the robustness of foundation models applied to lung cancer segmentation. Our analysis showed that whereas the models can exhibit similar volumetric segmentation accuracy, analysis of entropy as well as out-of-distribution robustness metrics, such as mis-detected tumor volume occupancy, provides a more comprehensive perspective on model performance. Our analysis shows that metrics combining OOD performance with task-specific metrics provide us valuable insights into understanding model behavior when deploying in clinical settings, where the ground truth labels are often unknown.

\clearpage

\acknowledgments 
 
This research was partially supported by NCI R01CA258821 and Memorial Sloan Kettering (MSK) Cancer Center Support Grant/Core Grant NCI P30CA008748.

\bibliography{references,references_ood} 
\bibliographystyle{spiebib} 

\end{document}